Final Report on:

ECCS/NSF Workshop on
Quantum, Molecular and High Performance Modeling and Simulation for Devices and Systems
(QMHP)

(Held April 16-17, 2007 at NSF Headquarters in Arlington, VA.)

Authors and Workshop Organizers:

Dr. Jonathan P. Dowling
*Co-Director, Hearne Institute for Theoretical Physics*
*Hearne Professor of Theoretical Physics, Department of Physics and Astronomy*
*Louisiana State University*
*202 Nicholson Hall*
*Baton Rouge, LA 70803*
*Email:* jdowling@lsu.edu

Dr. Gerhard Klimeck
*Technical Director, National Science Foundation Network for Computational Nanotechnology*
*Professor, Department of Electrical and Computer Engineering*
*Purdue University*
*Electrical Engineering Building, Room 313B*
*465 Northwestern Avenue*
*West Lafayette, IN 47907*
*Email:* gekco@purdue.edu

# 1 EXECUTIVE SUMMARY

**Scope and Goals:**

ECCS has identified a new thrust area in Quantum, Molecular and High Performance Modeling and Simulation for Devices and Systems (QMHP) in its core program. The main purpose of this thrust area is to capture scientific opportunities that result from new fundamental cross-cutting research involving three core research communities: (1) experts in modeling and simulation of electronic devices and systems; (2) high performance computing relevant to devices and systems; and (3) the quantum many-body principles relevant to devices and systems. ECCS is especially interested in learning how work in these areas could enable whole new classes of systems or devices, beyond what is already under development in existing mainstream research. The workshop helped identify technical areas that will enable fundamental breakthroughs in the future. Modeling and simulation in the electronics and optoelectronics areas, in general, have already resulted in important fundamental scientific understanding and advances in design and development of devices and systems. With the increasing emphasis on the next generation of devices and systems at the nano, micro, and macro scales and the interdisciplinary nature of the research, it is imperative that we explore new mathematical models and simulation techniques. This workshop report provides a better understanding of unmet emerging new opportunities to improve modeling, simulation and design techniques for hybrid devices and systems. The purpose the QMHP workshop was to define the field and strengthen the focus of the thrust. The ultimate goal of the workshop was to produce a final report (this document) that will be posted on the web so that the NSF can use it to guide reviewers and researchers in this field.

**Technical Merit:**
- Numerical approaches that can treat large enough systems (e.g., systems of around a million atoms at the nano-scale) under non-equilibrium conditions fully quantum mechanically including effects of dissipation and decoherence need to be developed and demonstrated. High Performance Computing (HPC) can enable the rapid solution of such systems.
- Develop a strategy for validation and verification. Numerical experiments must be repeatable, documented, and open just like physical experiments and theory.
- Integration of end-users into such fundamental developments is highly desirable, since their needs should determine the outcome of a highly available and usable modeling and simulation approach.

**Impact:**
We are currently in the midst of a second quantum revolution. The first quantum revolution gave us new rules that govern physical reality. The second quantum revolution will take these rules and use them to develop new technologies. In this review we discuss the principles upon which quantum technology is based and the tools required to develop it. A number of examples of research programs that could deliver quantum technologies in coming decades include: quantum information technology, quantum electromechanical systems, coherent quantum electronics, quantum optics, and coherent matter technology. These quantum technologies will underpin much of the technical and economical development of the World for the next 100 years. The fusion of the recent breakthroughs in high-performance computing with those in quantum technological materials development, as proposed in QMHP will accelerate the development of quantum technologies and their movement to the market place, as well as uncover new domains of scientific breakthroughs at the interface between the different subfields. We predict that quantum science and technologies will be one of the key enabling features of this new technological revolution, taking us beyond the singularity, for example, when Moore's Law for the exponential growth in computer processing speeds comes to an end in 10 years or so. A quantum Moore's Law must take over then, and the NSF's QMHP program will take us all into this new era.

**Recommendations:**

Recommendations for research directions are divided into two areas: quantum electronics and quantum photonics.

*Quantum Electronics:* In the electronics area, the recommended systems range from further studies of emergent devices replacing CMOS, especially those involving nano size and spin to explorations towards quantum systems employing entanglement and treating decoherence. The issues of methodology include formalism dealing with quantum transport, multi-scale modeling in time and in space, choice of typical systems for definitive computation, and algorithms which address the key issues of plug-and-play modeling tool, numerical stability, nonlinear equations for condensed matter systems, and inverse modeling for design, synthesis and characterization.

*Quantum Photonics:* In the photonics area, the primary goals are: (1) High-performance computing devices and systems; (2) modeling and simulation for photonic systems; and (3) quantum many-body principles devices and systems. The areas identified for research include: plasmonics, quantum computers, optimization and inversion, adaptive computing, quantum and classical boundary, materials aspect, and scale-up/integration. The issues include: quantum entanglement, nonlinear effects, phonon control, use of photonics crystals and metamaterials, as well as distributed and grid computing.

## 2  WORKSHOP OVERVIEW

The QMHP Workshop was held April 16-17, 2007 at NSF Headquarters in Arlington, VA, and supported by a grant from the NSF to Louisiana State University (LSU). The co-organizers of the workshop, and authors of this document, (Drs. Dowling and Klimeck), brought together experts in modeling and simulation of electronic and photonic devices and systems, with emphasis on high-performance computing and quantum many-body principles. The participants were loosely grouped into such experts in the fields of quantum electronics and quantum photonics. Details of the workshop may be found on the website: http://baton.phys.lsu.edu/~jdowling/qmhp/. The first day of the workshop was devoted to 20 minute focused presentations by the invitees, to coordinate the discussions. The second day of the workshop the invitees were broken down into quantum electronic and quantum photonic working groups, and then all reconvened to make joint recommendations on the program at the end of the day. This report is a summary of these presentations, discussions, and recommendations. We provide here below a program from the workshop, while more details (including copies of the presentations) can be found on the website above.

MONDAY 16 APRIL 2007

SESSION I (Chair: Gerhard Klimeck)

8:00AM OPENING REMARKS

Usha Varshney, Division Director
Division of Electrical, Communications and Cyber Systems (ECCS)
Directorate for Engineering, NSF

Paul Werbos, Program Director
Quantum, Molecular and High-Performance Modeling and Simulation for Devices and Systems
Division of Electrical, Communications and Cyber Systems (ECCS)
Directorate for Engineering, NSF

Jonathan Dowling
Gerhard Klimeck

8:20AM Avik Ghosh (U. Virginia) "Fundamental Issues in Nanoelectronic Modeling and Simulation"

8:40AM Kang Wang (UCLA) "Modeling Needs for Nanoarchitectonics and Nanosystems"

9:00AM Dragica Vasileska (ASU) "Process Variation and RF Performance Analysis in 10nm FinFET Using Contact Block Reduction Method"

9:20AM Yuli Lyanda-Geller (Purdue) "Controlling Dissipation and Decoherence in Nanodevices"

9:40AM Susan Coppersmith (U. Wisconsin) "Experiment, Simulation, and Theory for Developing High-Performance Quantum and Molecular Devices and Systems"

SESSION II (Chair: Gerhard Klimeck)

10:20AM Dmitri Nikonov (Intel) "Directions for Simulation of Beyond-CMOS Logic"

10:40AM Jerry Tersoff (IBM) "Low-Frequency Noise in Nanoscale Ballistic Transistors"

11:00AM Mark van der Schilfgarde (Arizona State U.) "The Quasiparticle Self-Consistent GW Approximation"

11:20AM Xuedong Hu (SUNY Buffalo) "Coupled Electron-Nuclear Spin Dynamics in Semiconductor Nanostructures"

11:40AM scheduled: Henry Everitt (AMCOM) "Quantum Information Science, as presented to the National Academy Decadal Study on AMO Physics"
11:40AM actual on short notice: Gerhard Klimeck, Purdue University, "*A Personal view on Challenges for Quantum System Modeling and Impact"*

SESSION III (Chair: Jonathan Dowling)

1:20PM Mikhail Lukin & Marko Loncar (Harvard) "Exploring New Interfaces Between Quantum Optics and Nanoscience"

1:40PM Marian Florescu (JPL) "Quantum Nonlinear Effects in Photonic Crystals: Applications to Information Processing and Thermal Radiation Management"

2:00PM Ting Shan Luk (Sandia) "Photonic Crystal Program at Sandia National Laboratories"

2:20PM Lu Sham (UCSD) "What Can be Done About the Decoherence of a Quantum System?"

2:40PM Eli Yablonovitch (UCLA) "The Application of New Algorithms to Solve Inverse Problems in Engineering"

SESSION IV (Chair: Jonathan Dowling)

3:20PM David Smith (Duke) "Electromagnetic Cloaking"

3:40PM Robert Trew (NCSU) "Thermal Emission in the Near-Field from Polar Semiconductors and the Prospects for Energy Conversion"

4:00PM Gary Brown (VA Tech) "Electromagnetic Scattering from Large Amplitude, Small Period Surface Roughness"

4:20PM Tae-Woo Lee (LSU) "Computer Aided Approaches for Nanophotonics Research"

4:40PM Joseph Eberly (U. Rochester) "First Steps and Results with Quantum Many-Body Math"

TUESDAY 17 APRIL 2007

8:00AM INTRO

8:20AM BREAKOUT SESSIONS I

A: Quantum Electronics (Chair: Gerhard Klimeck)

B: Quantum Photonics (Chair: Jonathan Dowling)

10:20AM BREAKOUT SESSIONS II

A: Quantum Electronics (Chair: Gerhard Klimeck)

B: Quantum Photonics (Chair: Jonathan Dowling)

1:30PM SUMMARY OF QE BREAKOUT SESSIONS TO ALL ATTENDEES (Klimeck)

2:15PM SUMMARY OF QP BREAKOUT SESSIONS TO ALL ATTENDEES (Dowling)

PANEL DISCUSSIONS

3:30PM PANEL DISCUSSION I — QUANTUM ELECTRONICS
(Chair: Gerhard Klimeck)
Panel Members: Kang Wang, Dragica Vasileska, Dmitri Nikonov, and Susan Coppersmith

4:15PM PANEL DISCUSSION II — QUANTUM PHOTONICS
(Chair: Jonathan Dowling)
Panel Members: Joseph Eberly, Marko Loncar, Lu Sham, David Smith, and Robert Trew

5:00PM FINAL WRAP UP

5:30PM ADJORN

# 3 INTRODUCTION

The purpose of the workshop and this report is to quantify and codify what is to be meant at the research thrust of QMHP. This document will help assist NSF program managers, NSF reviewers, and proposers to this thrust area. Technically, QMHP is a "thrust area" in the ECCS Division. It is listed as one of the thrust areas on the ECCS web site, but there is no special program announcement for QMHP or for anything else in the core of ECCS. (ECCS abolished program announcements back in 1994.) At present, to submit a proposal to QMHP, one simply submits a regular unsolicited proposal to ECCS. Beginning the title with "QMHP" and choosing the "organization unit" as "PCAN" makes it certain that it will go to QMHP and not to any other area of ECCS, but it is not required. When people submit proposals to QMHP or to any other core program in ECCS, in FASTLANE, the "program announcement" should simply be chosen as "GPG," the general program announcement for all regular core unsolicited proposals to NSF. Also, they should be submitted only during one of the two yearly "windows," in September and then in January. (All of this information is also on the NSF ECCS website.)

Due in part to the predilections of the two organizers, the invitees and hence the presentations and discussions naturally grouped themselves into the subtopics of "quantum electronics" and "quantum photonics" and the report will also be so subdivided. Within these two subdivisions, we will summarize the presentations given at the workshop, then summarize the breakout discussions, and finally give the recommendations and results of the workshop attendees. Individual attendees were solicited to write stand-alone comments, and these will be appended to the subtopics and identified as such.

# 4  QUANTUM ELECTRONICS

To set the theme for these discussions, we reproduce first the abstracts provided by each of the speakers in the quantum electronics talks that took place on Monday. (Complete copies of the talks are available on the above-mentioned website.) We will then synopsize the breakout panel discussion topics on Tuesday morning and then end with the panel discussion topics from the quantum electronics panel on Tuesday afternoon. We will end this section with a synopsis and recommendations from the individual invitees.

## 4.1  ABSTRACTS OF THE ELECTRONICS TALKS

Novel technologies have enabled the construction of new devices and materials at the nanometer scale that have the opportunity to introduce new computing, sensing, and actuation capabilities to mankind.  At that nanometer scale traditionally disjoint research endeavors meet:  device physicists, device engineers, material scientists, circuit engineers, mechanical engineers, and computational scientists / engineers.  At the atomic scale the distinction of material/device/circuit is vanishing and these formerly disjoint groups need to learn how to communicate their expertise to the other areas.  The quantum electronics group represented in this workshop spanned these different areas of expertise and enabled an identification of open questions, opportunities and anticipated impact.

*"Fundamental issues in nanoelectronic modeling and simulation,"* Avik Ghosh, University of Virginia.
The talk will focus on formal, computational and device level challenges for modeling and simulation of nanoelectronic devices and systems. *Formal* challenges involve quantum transport in the presence of strong many-body correlations (such as Coulomb Blockade), incoherent scattering (due to phonons, for example) and time-dependent effects (such as hysteretic switching and random telegraph noise). *Computational* challenges involve modeling surfaces and interfaces, where we need practical semi-empirical descriptions with transferable parameters to handle hybrid heterogeneous integrated systems. In addition, we need multiscaling and embedding techniques to merge these models with more detailed 'ab-initio' descriptions of chemically significant moieties. Finally, *Device* level challenges involve identifying fundamental limits of existing device concepts, such as molecular FETs, as well as exploring novel device operational principles. The talk will touch upon fundamental issues that arise in context of each challenge, and a potential route towards addressing them. I will then apply these ideas to a specific device architecture, namely, an ordered array of quantum dots grown on the surface of a nanoscale silicon transistor. All of the challenges identified above manifest themselves prominently in this geometry. Specifically, we will explore how the strongly correlated electrons in the nanoscale dots 'talk' to their weakly interacting counterparts in the macroscopic channel, how the interfacial electronic structure captures both long-ranged band correlations and short-ranged chemical correlations, and how the tunable coupling between the channel current with the charge, spin or vibronic degrees of freedom of the ordered dot arrays can lead to novel device paradigms such as scattering-based ultrasensitive detectors, memory elements and heat sinks.

*"Modeling Needs for Nanoarchitectonics and Nanosystems,"* Kang Wang, UCLA.  No abstract available.

*"Process Variation and RF Performance Analysis in 10nm FinFET Using Contact Block Reduction Method,"* Dragica Vasileska, Arizona State University. We utilized fully self-consistent quantum mechanical simulator based on Contact Block Reduction (CBR) method to optimize 10 nm FinFET device to meet ITRS requirements for High Performance (HP) logic technology devices. Fin width, gate oxide thickness, and doping profiles are chosen to have optimal and technologically realizable values. We find that the device on-current approaching the value projected by ITRS for HP devices can be obtained using conventional (Si) channel. However, getting a good subthreshold slope for specified on-current is challenging. Our simulation results also show that quantum nature of transport enhances the intrinsic switching speed of the device. The value of intrinsic switching speed projected by ITRS is achievable with our device. RF performance of the device has also been examined by extracting the unity current gain frequency. Sensitivity of the on-current, threshold voltage, DIBL, subthreshold slope etc. for the optimized device to process variation at room temperature have been investigated. Device performance has also been analyzed for slow process corner (high temperature, low operating voltage) to estimate the degradation in performance in terms of related parameters.

*"Controlling Dissipation and Decoherence in Nanodevices,"* Yuli Lyanda-Geller, Purdue University. The development of nanotechnology raises fundamental questions of quantum coherence, dephasing and dissipation in nanostructures. These concepts are vital for new devices and systems, for conventional computing and quantum information technology. I will (1) compare two major dissipation mechanisms in quantum dots, (2) show the role of system symmetry in dissipation, and (3) discuss the relation between decoherence and quasi-particle localization. As an example, I will describe mechanisms of nuclear spin decoherence in quantum dots, and show that relaxation times can be controlled via the Coulomb blockade charging effects, and changed by seven orders of magnitude by applying gate voltage. Finally, I will present my view on what are major scientific questions in this field.

*"Experiment, Simulation, and Theory for Developing High-Performance Quantum and Molecular Devices and Systems,"* Susan Coppersmith, University of Wisconsin, Madison. Simulation, theory, and experiment are all vital in the development of new materials, devices, and systems. This talk will discuss how close cooperation between researchers using all three approaches has enabled substantial new progress in the development of quantum dots in silicon/silicon-germanium heterostructures for quantum computing and spintronics applications.

*"Directions for Simulation of Beyond-CMOS Logic,"* Dmitri Nikonov, Intel Corporation. Continuous scaling of electronic transistors generated the need for modeling of nanostructured electronic devices. The result of it was recognition of a limit set by the laws of quantum physics. Non-charge-based logic on the molecular/atomic scale has to be explored in the hope of continued scaling. One promising option is spintronic devices, which are currently under experimental study in the academic community. The challenge for simulation is to achieve quantitative agreement and predictive power regarding such devices. Their advantage in power dissipation might arise from non-equilibrium operation and strongly correlated many-body states, which need to be theoretically designed. Non-charge-based systems will likely require non-traditional architectures, such as quantum cellular automata. High-performance simulation methods are required to model such systems with inclusion of both the device and circuit levels. Intel's efforts along these directions are briefly reviewed.

*"Low-frequency noise in nanoscale ballistic transistors,"* Jerry Tersoff, IBM Corporation. Low-frequency "1/f" electrical noise is ubiquitous, arising from thermally activated atomic-scale processes. As devices become smaller, they become increasingly noisy, and this represents a potentially crucial problem for nanoscale and molecular transistors in particular. Calculations for carbon nanotube ballistic transistors illustrate the novel scaling expected in this regime.

*"The Quasiparticle Self-Consistent GW Approximation,"* Mark van Schilfgaarde, Arizona State University. A new type of self-consistent scheme within the GW approximation is presented, which we call the quasiparticle self-consistent GW (QSGW) approximation. It is based on a kind of self-consistent perturbation theory, where the self-consistency is constructed to minimize the perturbation. QSGW describes optical properties in a wide range of materials rather well, including cases where the local-density and LDA-based GW approximations fail qualitatively. Self-consistency dramatically improves agreement with experiment, and is sometimes essential. QSGW avoids some formal and practical problems encountered in conventional self-consistent GW, which will be discussed. QSGW handles both itinerant and correlated electrons on an equal footing, in a true ab initio manner without any ambiguity about how a localized state is defined, or how double-counting terms should be subtracted. Thus QSGW combines advantages separately found in many kinds of ad hoc extensions to the LDA (e.g. LDA+U theory), in a simple and fully ab initio way. Weakly correlated materials such as Na and sp semiconductors are described with uniformly high accuracy; QSGW also reliably treats many aspects of correlated materials. Discrepancies with experiment are small and systematic, and can be explained in terms of the approximations made. Its consistently high accuracy make QSGW a versatile method that can reliably predict many kinds of materials properties in a unified framework, for example, critical energy-band parameters in GaAs and CuInSe2, and spin wave spectra in Fe and NiO. Also optical response can be calculated, as can scattering matrix elements such as the electron-phonon interaction (in principle); thus it can serve as an engine for true ab initio device design. As an illustration, QSGW output is coupled to a Monte Carlo Boltzmann Transport Equation solver to simulate a realistic GaAs MESFET without any empirical parameters.

*"Coupled Electron-Nuclear Spin Dynamics in Semiconductor Nanostructures,"* Xuedong Hu, SUNY Buffalo, Confined electron spins are promising candidates for qubit in solid state quantum information processing. Here we study coupled dynamics of electron and nuclear spins, particularly focusing on how this hyperfine interaction leads to decoherence in the electron spin. We also show that through gate control, the hyperfine interaction can be used to polarize the nuclear spins, thus reducing decoherence in the electron spin.

*"Quantum Information Science, as presented to the National Academy Decadal Study on AMO Physics,"* Henry Everitt (AMCOM). The National Academy of Sciences recently performed a decadal study on AMO physics, at which point quantum information science was discussed and plans for the future outlined. The presentation I gave during that study will be provided again so that workshop participants may understand some of the issues the NAS was addressing in the report, which has since appeared. The NAS was particularly interested in four issues, which will be addressed here:
 * What are the recent accomplishments, new opportunities, and compelling scientific questions?
 * What will be the impact on other scientific fields, emerging technology, and national needs?
 * What are the workforce, societal, and educational needs?
 * How might the US research enterprise realize its full potential?

Dr. Everitt from AMCOM had to cancel his attendance on very short notice. Prof. Klimeck took the open time slot to present his views on opportunities and challenges.
*"A Personal view on Challenges for Quantum System Modeling and Impact,"* Gerhard Klimeck, Purdue University. At the nanometer scale the concepts of device and material meet and a new device is really a new material and vice versa. While atomistic device representation is novel to device physicists who typically deal in effective mass models, the concept of finite devices that are not infinitely periodic is novel in the semiconductor materials modeling community. NEMO 3-D bridges the gap and enables electronic structure simulations of quantum dots, quantum wells, nanowires, and impurities. Electronic structure simulations of systems 21 million atoms have been demonstrated. To truly have impact on the research, experimental and educational efforts of the community, relevant tools must be put into the hands of experimentalists and educators. NEMO 3-D can engage both educators and advanced researchers, utilizing a single open source code. An educational version has been released on nanoHUB.org. Over 660 users ran over 4,100 simulations in the quantum dot lab in the past 12 months. The Network operates nanoHUB.org for Computational Nanotechnology (NCN). Over 21,100 users have used the nanoHUB in the 12 months leading to Feb. 2007. Over 5,100 users have launched over 174,000 simulations. We spend a lot of time and effort on diversity to enhance the representation of minority groups. This is of critical importance. But let us not forget that in science and engineering we have a true pipeline problem irrespective of diversity issues. Most graduate students in solid state device engineering and physics are from abroad. Unless we succeed in disseminating the research insights and inspire the next generation of scientists and engineers, we are doomed for failure in the American society to continue to be a leader in science and technology.

### 4.2   ELECTRONICS BREAKOUT DISCUSSIONS

On Tuesday morning Klimeck led the electronics breakout discussions. There were two three discussion segments:
1) Free brainstorm on ideas that come to mind when thinking about QMHP; ideas are just jotted down on various large post-it sheets, without in depth discussion and criticism.
2) The ideas are categorized and discussed.
3) Groups are assembled to report on the major categories. The reports are given below as a discussion result.

Five Major categories were identified and assigned to individuals to summarize:
1) Intellectual Challenge and gold standard benchmarks (G. Klimeck)
2) Big Ideas / Applications
    a. Systems
    b. Devices
    c. Replacing tomorrow's CMOS
    d. Dissipation and Decoherence

3) Methods
      a. Materials (M. van Schilfgaarde, A. Ghosh, J. Tersoff)
      b. Formalism (A. Ghosh, Y Lyanda-Geller)
      c. Multi Time / Multi-Length Scale Modeling (D. Vasileska)
      d. Algorithms (G. Klimeck)
   4) Dissemination / Engagement (G. Klimeck)
   5) Opportunities and Technological, Scientific, and Societal Impact (Wang, Klimeck)

## 4.2.1 Intellectual Challenges, Development of Gold Standards, and Deployment (Klimeck)

A fundamental challenge is the understanding and utilization of quantum phenomena under non-equilibrium conditions for realistically sized systems. Most of today's quantum mechanical approaches focus on properties at equilibrium of relatively small and idealized systems. However, real devices and applications operate out of equilibrium, redistribute electrons in phase space, involve decoherence and dissipation, and consist of systems of around a million atoms at the nano-scale. It is critical to comprehend that the non-equilibrium treatment with dissipation is of absolute criticality, while typical quantum mechanical approaches ignore it completely! Quantum mechanical systems without dissipation will repeat themselves after a certain period and cannot inject carriers on one end of the device and extract them at another end. Without those fundamental ingredients the wrong physics is modeled. Numerical approaches that can treat large enough systems under non-equilibrium conditions fully quantum mechanically including effects of dissipation need to be developed and demonstrated. High Performance Computing (HPC) can enable the rapid solution of such systems.

For Computation to truly become a third leg of Science, holding its own weight next to Experiment and Theory, it must develop its own strategy for validation and verification. Numerical experiments must be repeatable, documented, and open just like physical experiments and theory. They must be measured against agreed-upon standards before predictions into new realms are being made. The establishment of "gold standards" is a difficult one and may need the involvement of theory and experiment. For example even some basic bulk material properties of the most technology relevant material Silicon are unknown experimentally such as the strain behavior in various crystal directions. Different theories predict different behavior – well-qualified experiments are needed to develop verified and validated computational modeling approaches. New computational approaches are needed to treat certain classes of large problems, however, these new approaches should also be benchmarked rigorously against exactly solvable models to verify the approach.

The development of such algorithmic and numerical technology, however, is not good enough. These developments need to be deployed to end users that have real problems to solve. Integration of end-users into such fundamental developments is highly desirable, since their needs should determine the outcome of a highly available and usable modeling and simulation approach.

## 4.2.2 Big Ideas

### 4.2.2.1 Quantum Coherent Devices and Systems (Xu, Coppersmith, Sham)

**Device Proposals**: charge based, electron spin based, nuclear spin based, photon based, other quasiparticle or excitation based. New state-variables in CMOS; devices that exploit nonlocality.

The relentless pursuit of miniaturization of microelectronic devices has driven the feature size of the current generation devices below 50 nm. As dimensions of device elements get closer to atomic sizes of the order of nanometers, quantum mechanical features such as quantum tunneling, and level and charge quantization have become more prominent in undermining functionalities of classical electronic devices. Instead of struggling to overcome complications arising from the quantum phenomena, utilizing quantum-mechanical features, particularly quantum coherence, interference, nonlocality, and entanglement, to build novel quantum coherent devices should be increasingly explored. Fully coherent quantum computers exploit quantum parallelism can solve some problems much faster than classical computers, and they are being studied extensively at present. In addition, partially quantum coherent systems or small fully quantum coherent systems also deserve more attention in the hope that they may provide new functionalities or may perform classical tasks with reduced energy cost and at faster speed. Examples of applications where partial quantum coherent can lead to powerful new devices include quantum sensors, quantum lithography, and quantum repeaters. The degrees of freedom involved in these devices could be electron spin and/or charge, nuclear spin, photons, and various quasiparticles and collective excitations. Furthermore, exploring quantum mechanical features of classical devices such as MOSFETs could help reveal their full potential. In the exploration of quantum coherent devices, high performance computing will serve important roles in clarifying properties of complex quantum mechanical systems.

**Device characteristics**: Control of quantum coherence, decoherence, and dissipation and
**Device modeling**: System-level quantum modeling

In order to take full advantage of quantum mechanical properties of a quantum coherent device, the relevant quantum coherence of the system must be protected. It is important to identify all the external degrees of freedom with which the quantum device interacts, to clarify the specific channels for decoherence, and to devise approaches to eliminate, suppress, or control decoherence (including pure dephasing and energy-dissipative relaxation) processes. It is also important to perform system-wide simulations that account for all the (often classical) control knobs together with the quantum coherent components themselves. When the spatial dimensions of such devices are sufficiently small, the controlling gates/means are not fully independent and often interfere with each other. Furthermore, it is also inevitable that disorder, defects, and fluctuations are present in nanoscale devices. It would thus be crucial to simulate and analyze the degree of control one can achieve in the presence of such correlations, in addition to the potential decoherence from imperfections. Such detailed exploration of the boundary of classical and quantum regimes could have profound impact on our understanding of quantum mechanics and our understanding of the potentials and limitations of quantum coherent devices.

**Device i/o**: Quantum measurement

For a practical quantum coherent device, reliable and rapid measurement is required. Since measurement inevitably involves the transfer of information about the quantum state to the measuring device, it is crucial to explore how it can be done properly without interfering with the coherent operations of a particular device.

### 4.2.2.2 Systems (Xu)

Over 30 years ago Philip Anderson wrote a famous paper called "More is Different," in which he argued that fundamentally physical phenomena arise when many degrees of freedom are coupled together, even if the behavior of each of the individual degrees of freedom is well-characterized and understood (Science 177, 393-396 (1972)). This insight, which has been amply verified in the context of systems in condensed matter physics, promises to be equally important in the design of complex electronic and photonic systems.

One key question is how to optimize robustness. Robustness itself can be interpreted in different ways. One way is to be able to design systems that continue to work when the device yield in the initial fabrication is less than 100%, but in which the devices that work initially continue to work as the system is operated. Another type of robustness involves developing systems whose performance is minimally degraded when various components fail, or when the system is subject to noise. The question of how to develop system designs, which are robust to noise and to variability in the individual components, is becoming increasing compelling because of the continual shrinking of devices, resulting in larger relative fluctuations and in greater variability in the individual device characteristics. Developing the fundamental principles underlying robust design could benefit from investigating the architecture of complex biological systems, which are composed of components that exhibit substantial variability and yet perform sophisticated functions robustly.

Two other capabilities that would have enormous impact on future generations of devices are being able to manage heat more effectively and being able to minimize power requirements for an overall system with components of given individual characteristics. As such, the clarification of electron-phonon coupling in nanostructures with reduced dimensionality and material inhomogeneity is of crucial importance. Furthermore, exploring possible ways to control electron phonon coupling in terms of its strength and possible directionality could result in more efficient heat management and ultimately efficient recycling of thermal energy.

Another issue where an effective solution could revolutionize system integration is the search for the most efficient, reliable, and cost-effective ways of integrating different types of components. Examples include the interface between electronic and optical components, and between mechanical and electronic components.

### 4.2.2.3  Devices (Vasileska)

In the field of electronic devices we have two categories: (a) existing technology, and (b) emerging technology.

One way is to pursue constant in existing technologies in CMOS scaling and the other one is to look for alternative device geometries and/or new materials. In all these devices, unintentional dopants/defects and the non-locality of the carrier transport will play significant role on device operation. Therefore, 3D simulators are needed for modeling complex 3D structures made of arbitrary materials and with arbitrary crystallographic directions that will be quantum mechanical in nature to capture quantum-mechanical size quantization and quantum coherence effects.

Regarding emerging technologies, in which we can include non-charge-based devices, nuclear-spin based devices and devices used in quantum computing, simulation tools are needed that can properly capture, for example, the spin carrier transport which is not as yet developed at that level of sophistication as the carrier-based transport. In both, the existing and the emerging technology, modeling of the contacts will be crucial for proper understanding of the device operation. In fact, one cannot model an isolated device, but it must model the device + part of the reservoir. Another issue is the proper bandstructure incorporation, as at these scales devices will behave as macro-molecular systems. Hence, a transport + band-structure kernel is needed in the nanoscale device modeling.

### 4.2.2.4 Challenges for replacing tomorrow's CMOS (Nikonov)

Challenge: find the best option for the nanoelectronic transistor
- predict a nanoelectronic device dramatically better than CMOS
- prove when quantum transport simulation is necessary (i.e. drift-diffusion-Schrödinger-Poisson breaks down)

Challenge: choose best non-charge-based device
- tools for simple simulation of non-charge-based devices
- predict which options are dramatically better than electronics

Challenge: simulation capability to enable experimental demonstration of a practical spintronic device
- general spintronic simulator, handling most of proposed devices
- achieve predictive power by close interaction with experimentalists
- to filter device options, not to discover new phenomena

Challenge: can one lower power dissipation via operation out of thermal equilibrium?
- practical device scheme with non-equilibrium operation
- prove advantage of spin logic vs. electronic logic

Challenge: what are the architectures of choice for beyond CMOS devices
- simulator with device-circuit link for beyond CMOS devices
- which architecture brings out the best performance of non-charge-based devices
- any application-specific architecture for beyond CMOS that is dramatically better than CMOS

Challenge: how can HPC be made useful for real engineering?
- Need to perform simulations rapidly with the goal to help experiment
- Rapid turn-around time
- Need almost interactive response – as opposed to week-long numerical experiments

### 4.2.2.5 Dissipation and Decoherence (Lyanda-Geller)

For quantum devices on nanoscale, control of dissipation and decoherence becomes especially important. Modeling of nanodevices requires identification of principal mechanisms of dissipation and decoherence. Unlike dissipation in bulk materials, special role in dissipation in quantum dots and similar structures belongs to relaxational dissipation, when energy levels oscillate with frequency and level population adiabatically follows excitation, becoming non-equilibrium in the absence of interlevel transition, so that population relaxation leads to absorption of energy. Dissipation and decoherence in devices strongly depend on the symmetry of wavefunctions, which define strength of excitation and relaxation processes.

It has long been appreciated that there is an indispensable connection between dissipation and decoherence, with most striking examples being (i) the insulating properties of two-dimensional conductors in absence of decoherence and metallicity in the presence of decoherence, and (ii) absence of absorption in resonant transitions between two discrete levels in the regime of Rabi oscillations, with absorption of energy appearing with dephasing. Developing methods of control of decoherence is crucial for quantum information technology, for coherent electronic devices and for non-charge based devices, involving electron or nuclear spin. Critical issues in investigation of dissipation and decoherence are: account of dynamics and correlations in the bath, and correlations between processes in the device and in the bath, such as dependence of coupling between quantum dot and contacts on the number of particles in the quantum dot. It is critical to understand the role of many-body localization, and the role of localization in Fock space of many-body eigenstates for dissipation and decoherence.

Modeling of nanodevices requires an appropriate formalism. To investigate dissipation, decoherence and transport in quantum charge and non-charge based devices, it is insufficient to rely on Kubo formula, and using non-equilibrium Green functions and quantum kinetic (Boltzmann-type) equation is important, which would account for effects of disorder, interactions and correlations.

### 4.2.3 Methods
#### 4.2.3.1 Formalism (Ghosh and Lyanda-Geller)

Quantum and nanoscale systems and devices span a wide range of length, time and energy-scales. Physical processes at interfaces and contacts, where these scales come together, are often crucial. The issues are not just computational, but in fact, fundamental ˆ at the level of basic notions and equations themselves. How would one describe quantum transport through a small channel or quantum dot, including many-body interactions, disorder, correlations between charge/spin carriers in the dot/channel and those in contacts? What defines dissipation and decoherence in such systems, and how would one introduce level broadening rigorously? What is the role of many-body localization, and localization in Fock space of many-body eigenstates for transport properties, dissipation and decoherence? For instance, quantum kinetic equations are quite successful in describing nonlinear response characteristics of many systems. Can one use this approach to describe strongly interacting nanosystems far from equilibrium, or are fundamentally different non-perturbative approaches are needed? If so, what is the dynamics at the nano-micro interface, where strongly correlated elements couple with weakly correlated counterparts? What is the role of coherence and localization of quasi-particles at the system level? Could one borrow concepts from other disciplines (such as quantum optics) to describe some of the issues (such as entanglement) in quantum transport paradigms?

#### 4.2.3.2 Multi-Time / Multi-Length Scale Modeling

There are several ways of interpreting multi-scale transport. When we talk of length-scales, we usually mean starting from the Schrödinger equation and going up to the circuit level if we consider modeling of devices. This is becoming very important even in conventional nano-scale devices as quantum interference, non-locality of the transport and quantum size-effects play significant role. There are problems that can be also multi-time scale and multi-length scale at the same time. Such is the problem of heating of devices fabricated in the SOI technology where simulation tools are needed to capture physical phenomena starting from the circuit level (determination of the hot spots in the circuit) and moving down to the device level where we have multi-time scale due to the difference in the energy-relaxation times of the electrons, optical phonons, acoustic phonons and the underlying $SiO_2$ level (the heath bath).

#### 4.2.3.3 Materials (Ghosh, Schilfgaarde)

In the modern nanoscale and computing devices the chemistry, bandstructure and transport are strongly linked. Aside from the formal challenges of merging classical and quantum transport in devices spanning various length, time and energy scales, there is also the practical issue of identifying the correct electronic structure input to suitable transport solvers. Broadly speaking there exist a hierarchy of methods to model bandstructure. At one extreme are ab-initio methods, which contain explicit information about wave functions and potentials. They can be predictive, but are however limited in scale. At the other end are model Hamiltonians, whose matrix elements are parameterized to fit key materials parameters, and may not be easily transferable between structures. A critical issue in modeling practical devices is that complex devices often incorporate very different types of materials, such as organic and inorganic molecules, which pose a significant challenge to the modeler. One path is to develop multi-scale simulators, - approaches that couple more accurate methods in small, critical regions to simpler, more efficient ones that can be applied to larger scale. Another is to evolve the ability to create new electronic structure approaches that are more broadly applicable across boundaries and interfaces. Some combination of these strategies will be needed to realize the objective of scaling our results from the device level all the way up to a systems level.

With the smaller size of devices approaching nanoscale, effective mass methods commonly employed to describe charge carrier spectra, field effects and impurities in bulk systems no longer suffice. Atomistic approaches are needed to perform quantum simulations in practical devices. The latter are bound to have at least 1 million atoms; that is the capability that such computational tools should target. Furthermore, for investigation of the device functionalities related to changes of shape of devices by electrostatic gates, for study of effects of randomness due to fluctuations in alloyed materials, and effects of variability atomistic calculations can be effectively supplemented/combined with Random Matrix Theory methods. Random Matrix theory uses symmetries of systems, the corresponding energy level repulsion and statistical distribution of matrix elements of interactions for calculations of mesoscopic systems and elucidation of universal properties of devices and systems.

Accuracy is a major issue. New, efficient strategies are needed to supply information to adequately characterize impurities and defects, alloys, unfamiliar materials, interfaces between dissimilar materials, lattice deformations, and so on. The structure of the interfaces within the device are not known (particularly when molecules or carbon nanotubes are involved), and cannot be readily determined experimentally. Also, thermal fluctuations in structure and charge state lead to significant noise in micron-scale devices, and noise increases with decreasing size. Optimizing for structural stability and robustness against thermal fluctuations or diffusion may become as important as optimizing the electronic response of the device. Thus, numerical techniques need to be developed that can adequately address structural, electronic, and thermal properties. In addition, we need ways to describe device-to-device variability. Atomistic techniques may need to be combined with other symmetry approaches, such as random matrix techniques. While *ab-initio* methods can play a crucial role in addressing many of these issues, challenges to a practical realization need to be addressed. For example, the most widely used *ab-initio* methods have well known limitations to their accuracy, which often renders them unsuitable to describe key parameters. Moreover, such methods usually stay within the framework of a single-particle picture, which are inadequate for some important many-body correlation effects, particularly relevant for quantum transport. New frameworks need to be developed to overcome such limitations.

### 4.2.3.4 Algorithms (G. Klimeck)

The panel identified 3 general critical issues that need to be addressed in subsequent research work and several direct suggestion of needed algorithm work.

1) plug-n-play family of modeling tools – traditional software development at the University level in the are of quantum electronics serves as a supplement to the outcome of a PhD thesis. Typically the primary outcome of a PhD project is some new facet of specialized knowledge documented in a peer reviewed journal. The actual software that may have been used in the work is not really subject to review, documentation, and release. In almost all cases the software dies with the departure of the student. In that process many wheels get reinvented all the time, progress is slow and the software is inefficient, buggy, undocumented, and non-interactive with other systems. This is a systemic problem in the existing funding process and academic value system. NSF can strongly influence this by demanding its projects to deliver and disseminate the new software, to make it part of the peer review process, and to truly elevate computation to a third leg of science.

2) Numerical (in-) stability: Typical software designs and implementation are not modular enough (see argument above) to test and validate alternative solution methods on the same problem setting. This handicap severely hampers the testing, validation, and verification with alternative algorithms and the analysis of algorithm stability issues. The important thing to keep in mind here is that the new algorithms are not the "end", but just a means to the end to solve real problems. Computational scientists who develop new algorithms must be embedded in the physical problem and integrate their new algorithms in real software solving real problems, not toy or sample problems.

3) Algorithms for non-linear equations applied to condensed matter: Other disciplines have developed and implemented large sets of algorithms that may prove to be very useful to solve problems in condensed matter physics. Future announcements should encourage the cross fertilization of formerly disjoint computational efforts to help advance the computational science in condensed matter electronics.

4) Design, Synthesis, and Characterization through inverse modeling: Most modeling and simulation approaches concern themselves with accurate representation of some physical system and the derivation of its physical properties or technical performance. This is the baseline for the work of an engineer who will now take that knowledge and approach and will try to tweak and modify it so it actually delivers desirable outputs and performances. That means that typical engineering problems start from a desired performance and most of the time is spent to solve the "inverse problem" to find the physical properties, dimensions, material choices etc, that result in the desired performance. Automated approaches integrated into realistic, not toy model-based, problems need to be sponsored by the NSF. Such automated approaches will help the design and synthesis but they can also help the device and system characterization, since physical measurements are often hard or impossible and software can be used to characterize devices and components.

The sections on Systems, Materials, and Formalism laid out some of the intellectual challenges the field is facing. New or augmented algorithms are needed to advance the science or overcome significant obstacles. Some of the identified issues are:

1) Circuit simulation for non-traditional architectures: today's semiconductor circuits contain about 1 billion transistors. These systems are designed by a hierarchical system / circuit approach. Alternative architectures are currently suggested and explored at the few devices level. Overall circuit level analysis must be performed in order to help guide and validate the individual device work. What if in the wildest dreams the individual devices work as well as imagined, but put together in circuits they cannot deliver more than what we have today? These are validations system level circuit analysis needs to deliver.

2) Non-equilibrium Green's functions (NEGF) have been identified as one of the primary formalism to solve the quantum effect dominated non-equilibrium transport problems at the atomic scale. For some problem settings this approach poses daunting computational burdens for today's algorithms. The problem at hand is not the traditional eigenvalue problem or the linear system solution, but selective elements (typically all the diagonals and super-diagonals) of the inverse of the system matrix need to be computed. New, fast algorithms are needed for this problem solution.

3) NEGF can treat the quantum mechanics of a single carrier in an atomistic basis representation, treating interactions with other carriers through self-energies. This is appropriate for systems with a relatively large number of carriers. If the number of carriers are small, large effects arise due to true few particle-particle interactions and correlations. No generalized theory is even available today (see formalism section). The need for algorithmic work on eigenvalue solutions is expected to have a strong impact on the treatment of these strongly correlated systems.

### 4.2.4 Dissemination and Engagement (G. Klimeck)

The growth in knowledge in all fields of science is so tremendously large that the knowledge dissemination and advancement faces fundamental challenges with respect to 2 very different population groups: people inside and outside the science / engineering community.

1) Dissemination inside the Science and Engineering communities. Traditionally computational scientists are very decoupled from experimentalists and theorists. What is exciting and new to a computational scientists is typically too complicated and too computationally intensive, or too inaccessible to an experimentalist. And what is interesting and essential to an experimentalist is often an old and boring hat for a computational scientists. Even for computational scientists it is difficult to cross borders to a new area, such as device modeling towards materials modeling, as codes are hard to access, maintain, and use. NSF must encourage the interaction of these typically disjoint groups and must demand that the computational work be made accessible to experimentalists and the community at large. There is currently no or little formalized and funded encouragement by NSF to finalize and disseminate software developments in research efforts.

   One successful example where experimentalists, educators, computational scientists, and students can use research and educational software very easily is nanoHUB.org developed and hosted by the Network for Computational Nanotechnology. The nanoHUB is now serving over 25,000 annual users with nanotechnology simulations, tutorials, seminars, web meetings, and collaborations. New middleware, application development environments, and portal technology was developed to serve real users and this technology is now being used to create new HUBs. NSF should encourage the use of similarly proven technologies for dissemination and engagement, rather than the creation of new future Cyber infrastructures as research efforts, that will take another few years to develop and yet more years to serve real users. Cyber infrastructure must be usable infrastructure that serves real users immediately. The real usage guides the evolution of that infrastructure.

2) Even if the wildest device and systems dreams were to come true in a very short time frame, science and engineering are facing a tremendous challenge in the US as well as many other technologically highly developed countries: children in these countries do not grow up wanting to be engineers and scientists! Making research tools, research seminars, or introductory lectures available to K-12 groups cannot be the right way to address these groups. Specialized efforts that focus on the integration of fundamental engineering and science knowledge into the existing curricula, motivated by today's science problems may be a great opportunity. This can be a true service opportunity of scientists and engineers who are gifted teachers, who can distill essential knowledge in math and physics into a K-12 curriculum.

### 4.2.5 Opportunities Technological, Scientific, and Societal Impact (Wang, Nikonov)

The QMHP thrust will result in commercialization of a new generation of devices for computing, sensing, information storage and processing, which can only be enabled by the simulation and design methods identified in the present workshop. The impact of the thrust will occur in the technological, scientific, and societal aspects.

**Technological impact**. The field that is most significantly dependent on the implementation of the nanoscale and molecular scale devices (electronic, nanomagnetic, spintronic, optoelectronic, nanomechanical, etc.) is computing. For 35 years to date, the progress of computing devices according to Moore's law occurred by their dimensional scaling. Semiconductor technology developers encounter prohibitive obstacles in continuing the trend. The only possibility to sustain the Moore' law for next few decades seems to come from dramatically novel devices such as discussed in this paragraph. Such devices are necessary for higher performance and lower power computers.

Benefits are not limited to general purpose computing. It is likely that the considered quantum and molecular devices will manifest the strongest advantage in non-traditional architectures and specific applications, such as image recognition, data sorting, digital signal processing.

New heterogeneous integrated Nanosystems for information (sensing, energy management, low dissipation information processing, health and society, self sustaining.) will emerge with the major NNI investment.

Another branch of this thrust in the thermal management and the hope of harvesting of the thermal energy of the environment in the nanoscale heat engines to provide energy to mobile or remotely controlled information devices.

**Scientific impact**. Pursuit of the QMHP program will require novel approaches to fundamental physics. Here we stress that, physicist will need to tackle switching and variable states of condensed matter and quantum object, unlike mostly spatially uniform and stationary states of matter they are used to study today.

The understanding of variability and power dissipation, which are the common salient features for all nanometer scale devices with high scales of integration, will enable the new robust design and operation for the next generations of Nanosystems.

The simulations tools as well as theory developed will accelerate the development of high performance simulation for nanosystems by providing new understanding of the interactions of nanocomponents (including physical, chemical and bio building blocks) and their interfaces amongst different dissimilar nanocomponents for heterogeneous integration. Therefore, seamless modeling and simulation from nanocomponents to nanosystems are needed. The general scope of quantum information simulations should cover a broad scope of nanoscale quantum systems with modules to be seamlessly integrated for users.

**Societal impact**. Deployment of systems based on the new device will initiate profound changes in the society. It took the humankind approximately 50 years to reach the mark of 1 billion computer users in the world. We should reach the Second Billion of computer users and connect them over the internet in a much shorter time. The fates of growth of global economy, proliferation of information freedom and democracy in the world, increase of living standards across the world hinge on the success of information technology.

Achievement of a qualitatively new level of application specific computing (mentioned above) will enable unprecedented capabilities in security, defense, and service economy.

The quantum information we learned may provide new sensing and measurements in nanoscale (atomic and molecular scales): new quantum sensors and measurement techniques to go beyond what we have today, particularly in the quantum systems may help many facet of social life. For example, the quantum fluctuations in the nanoscale, which may correlated, may be treated as signal for metrology as contrast with the microscale, where these fluctuations are considered as "noise". New theory of quantum metrology and correlated effects is needed to help develop new experimental quantum based instrumentation. Applications may include those in information systems as well as health systems, e.g., improvement of MRI.

A strategy for the seamless integration of different modules of high performance computing should be developed. For different modules, they should be able to help systems designers to improve their effectiveness in design and to increase of the productivity by orders of magnitude. The latter will improve all facet of productivity in the society.

### 4.3   Electronics Panel Discussion

Panel Members: Kang Wang, Dragica Vasileska, Dmitri Nikonov, and Susan Coppersmith, Panel

Chair: Gerhard Klimeck

The process of the panel was such that each panelist was given 3-5 minutes to make their initial statements and the floor was opened up for discussion. The paragraphs below highlight the keywords of the individual statements and the ensuing discussions.

*Kang Wang:*

What will be the anticipated impact of this program 10 years from now? What would you anticipate? Do not have an answer for these questions.

*Susan Coppersmith:*

Thinking of QMHP as a program: Why is this not more of the same, but really be different? Need to put the material into context – impact here is the direct relevance to technology. Foresee big change in electronic devices in the future. How do we keep this in the real footing and how do we tie this to real technology? Has to couple to the real device world – self consistency of the model to the real world!

Needs to be a critical new idea! Have not heard a new idea in the discussion so far

Need to really form the sense of a community.

*Dragica Vasileska:*

Wants to be a bit more specific! What are the current problems? What are the relevant problems? What are the modeling tools that are needed for the end of the road map and beyond?

*Dmitri Nikonov:*

Need to consider quantum device and molecular computing as a target.

No one really touched on high performance computing in the previous discussions. Usually think of HPC of a person that publishes something that cannot be duplicated. HPC needs to be connected to real solutions that are fast and accessible.

Discussions:

Kang Wang: HPC means today access to remote systems – need to see ubiquitous HPC – in educational and research institutions.

Dmitri Nikonov: HPC in the context in this thrust – in order to achieve the goals the researchers need to adopt much more efficient way to compute, formulate and distribute. Not necessary to use a million of computers to solve the problems – need to get the optimization done quickly.

Gerhard Klimeck - HPC: needs to be useful to engineers to turn solutions that normally require days of compute time into minutes. If such HPC simulations are readily available cycles and user interfaces) then engineers will use them.

Eli Yablonovitch: needs a certified software genius to make things work in software. There are no turn key solutions that will work for a large group.

Kang Wang: need HPC for fundamental calculations – has not seen it applied to immediate realistic problems

Summary of the main part of the discussion:

In the past HPC has not been particularly helpful for engineering work. Efforts under this program in HPC should be closely coupled and integrated into the effort to have any impact. Some output of this thrust should go to software development and software engineering. => broader impact. What kind of software development is needed to speed up the development?

# 5   QUANTUM PHOTONICS

To set the theme for these discussions, we reproduce first the abstracts provided by each of the speakers in the quantum photonics talks that took place on Monday. (Complete copies of the talks are available on the above-mentioned website.) We will then synopsize the breakout panel discussion topics on Tuesday morning and then end with the panel discussion topics from the quantum photonics panel on Tuesday afternoon. We will end this section with a synopsis and recommendations from the individual invitees.

## 5.1   ABSTRACTS OF THE PHOTONICS TALKS

Many of the talks, and hence the discussions, focused on high-performance simulation, modeling, design, and experiment of complicate electromagnetic systems. Examples are the electromagnetic properties of metamaterials, photonic crystals, plasmonics, and the computational difficulty involved in modeling the field-matter interactions, such as the nonlinear inversion problem — what design gives the desired electromagnetic response? In the case that the field-matter interactions are nonlinear in nature, the inversion can be formidable and new high-performance computing algorithms and techniques are needed. An additional layer of computational complexity occurs when the matter or field (or both) need to be treated quantum mechanically, for example in the design of quantum optical systems for quantum information processing. Here we run into the paradox, for example, that in order to design a large-scale all optical quantum computer, one may need a large-scale quantum computer. The analog is that in order to design the next generation of computer chip, one must exploit a supercomputer that employs the previous generation of computer chips. Bootstrapping in the quantum domain will become similarly important. The understanding of light matter interaction at the quantum and nanoscale will be come critical and high-performance computing must address this need. Finally the focus will be on optoelectronic quantum computers, their design and their uses.

*"Exploring New Interfaces Between Quantum Optics and Nanoscience,"* Mikhail Lukin & Marko Lonco, Harvard University. Several examples from the emerging interface of quantum optics, photonics and nanoscience will be discussed. We first describe and demonstrate a technique that makes use of coherent manipulation of a single electron spin and individual nuclear spins in its high purity diamond lattice to create a controllable quantum system (quantum register) composed of a few qubits. Specifically, we show that the electron spin can be used as a sensitive local magnetic probe that allows for a remarkable degree of control over individual nuclear spins. We next describe our recent work on developing novel systems for single photon quantum control based on combining ideas of cavity QED with nanowire surface plasmons. Specifically, we will describe realization of strong coupling between such plasmons with CdSe quantum dot emitters. As an outlook we will discuss how these systems can be used for potential applications ranging from scalable quantum information systems to novel sensors and single photon transistors.

*"Quantum Nonlinear Effects In Photonic Crystals: Applications To Information Processing And Thermal Radiation Management,"* Marian Florescu, Louisiana State University and Jet Propulsion Laboratory. Photonic crystals constitute a new class of dielectric materials in which the basic electromagnetic interaction is controllably altered over certain frequency and length

scales. They promise a wealth of new devices that could satisfy the demand for ever-faster computers and optical communications. The synergetic interplay between material science and high-performance modeling and simulations is the major driving force in the field of photonic crystals. In this talk, I will review some of the novel phenomena in the quantum optics of photonic crystals and I will present our work on all-optical switching near a photonic band edge, thermal radiation control using photonic crystals, and implementation of deterministic photon sources in photonic crystal architectures.

*"Photonic Crystal Program At Sandia National Laboratories,"* T.S. Luk, I. El-Kady, R.A. Ellis, J.D. Williams, W.W. Chow, Sandia National Laboratories, Albuquerque. The ability of a photonic crystal to mold and control light forcing thermal energy to emit in a relatively narrow spectral band has intrigued scientists and engineers in optics and physics. In particular, there is interest in driving photonic crystal emitters out of equilibrium. In this way, Planck's law will no longer limit the spectral brightness of the emission. Indeed, a quantum optical model of an ensemble of two level emitters suggested that a nonequilibrium condition could be achieved even when the heat source is thermal in nature. This is due to the fact that the temperature of the emitters is not necessarily the same as the photonic crystal itself. This result has tremendous implications for thermophotovoltaics, efficient lightning and IR scene generator applications. Sandia has a long history of photonic crystal research. In this presentation, we will discuss the direction and focus of Sandia photonic crystal program. This will include modeling, fabrication and characterization. In addition, we will touch on phononic bandgap materials, the acoustic analog of photonic crystal.

*"What Can Be Done About The Decoherence Of A Quantum System?"* Lu J. Sham, University of California San Diego. The superiority of a quantum machine is rooted in the coherent superposition of its states. This talk will discuss two methods of reducing decoherence: one restores the lost coherence and the other a nascent proposal to avoid decoherence. (1) The coherence dynamics is modeled by a two-level system in an interacting spin bath. Based on a quantum solution of the dynamics, a method is proposed to recover from the pure decoherence. (2) Adiabatic quantum computation can avoid decoherence by keeping the system in the ground state by changing the fields. We discuss whether the density functional theory can be used to model the adiabatic change of the ground state.

*"The Application of New Algorithms to Solve Inverse Problems in Engineering,"* Eli Yablonovitch, Department of Electrical Engineering, UCLA. The solution of Maxwell's Equations is central to many areas of technology. I will give the following real-world business examples in which a more accurate Electromagnetic computation redounds right down to the bottom line of a business: 1. Cell phone Internal Antenna design; 2. Nano-photonic circuits in Silicon-on-Insulator wafers; 3. Design of photo-masks for Photolithography. Simulation alone is insufficient in all these real life examples. Ideally the computation should encompass an inverse design algorithm that actually solves the engineering challenge.

*"Electromagnetic Cloaking,"* David R. Smith, Duke University. Over the past five years, artificial materials—or metamaterials—have rapidly emerged as a means of obtaining unusual electromagnetic response. While the range of values that the electromagnetic constitutive parameters can exhibit in metamaterials does not necessary exceed what can be found in conventional materials, the ability to engineer the electromagnetic response over a tremendously broad frequency range far surpasses what can be achieved in conventional materials. This design flexibility has enabled the development of a suite of new electromagnetic materials, from media with negative refractive index—in which the electric permittivity and the magnetic permeability must be negative over the same frequency interval—to impedance matched gradient index structures. Most recently, we have suggested an entirely new approach to electromagnetic design; termed transformation optics, in which an optical device is created by the application of a desired coordinate transformation to the constitutive parameters. As a first example of this method, we describe both the theoretical recipe as well as an experimental confirmation for an "invisibility cloak," which has the property that it sweeps electromagnetic waves of a particular bandwidth around a concealment volume, neither reflecting radiation nor casting a shadow. The cloak parameters determined by the coordinate transformation are generally quite complex, being anisotropic and exhibiting significant inhomogeneity in both the permittivity and permeability tensors. Though the cloak structure is complicated, the recent demonstrations point out the further enabling possibilities of metamaterials.

*"Thermal Emission in the Near-Field from Polar Semiconductors and the Prospects for Energy Conversion,"* R.J. Trew, V.N. Sokolov, B.D. Kong, and K.W. Kim, North Carolina State University, Raleigh. It has been demonstrated that in the near field the emission spectra of thermal radiation from polar semiconductors exists in narrow bandwidth evanescent modes at THz frequencies. Quasimonochromatic thermal emission exists in material sources that support surface phonon polaritons. It is shown that polar semiconductors, such as InP, GaAs, GaN, SiC, and α-Al2O3 (sapphire), all exhibit quasimonochromatic thermal emission characterized by strong peaks of evanescent modes at well-defined frequencies in the near-field that correspond to the appropriate peaks in the density of states for surface phonon polaritons. Materials with lower polariton frequencies (e.g., InP and GaAs) demonstrate a higher peak spectral energy density compared to those with higher frequencies (e.g., SiC). The energy density stored in the evanescent peaks close to the surface is estimated to be many orders of magnitude larger than that in the blackbody radiation. Conversion of the near-field evanescent mode to a traveling wave suggests the potential for a novel energy source that can convert thermal energy directly to dc current.

*"Electromagnetic Scattering from Large Amplitude, Small Period Surface Roughness,"* Gary S. Brown, Virginia Polytechnic Institute & State University, Blacksburg. Our understanding of scattering from rough surfaces rests primarily upon asymptotic analytical models and numerical results. The analytical results have been found to be much more accurate than was originally predicted. The numerical results have shown this to be true and have also extended our domain of understanding to situations where the asymptotic analytical results breakdown. Boundary perturbation theory is accurate when the surface features are small compared to the probing electromagnetic (em) wavelength. The Kirchhoff approximation is accurate when the roughness height and period are large compared to the electromagnetic wavelength. An appropriate combination of these two theories covers surfaces, which are a mixture of small and large structure. The intermediate range where the surface height and the period are comparable to the em wavelength can be dealt with using numerical techniques. The one range of surface roughness that has not been addressed to date is that for which the roughness height is large while the surface period is small compared to the em wavelength. If the surface roughness period becomes sufficiently small compared to a wavelength, one might expect that the surface will become specular in it scattering characteristics. That is, from a scattering point of view, the surface effectively becomes flat. The purpose of our research is to investigate this limiting behavior in order to understand the scattering from such surfaces. Our goal is particularly focused on finding what happens to the currents on and the scattered fields from randomly rough surfaces comprising surface periods that are smaller than the incident em wavelength. Our approach focuses on the Magnetic Field Integral Equation (MFIE) for the surface current induced on a one-dimensionally rough, perfectly conducting surface. Unfortunately, in its primitive form, the MFIE is not particularly amenable to numerical solution for scattering by an extended surface. We therefore use a preconditioning technique that we developed in 1996 for summing an infinite number of multiple scatterings on the surface prior to an iterative solution. We demonstrate the robustness of the method by applying it to a surface containing a sinusoidal roughness. We show the rather unique behavior of the surface currents for both TE and TM polarizations once the period of the surface becomes less that one-half the incident electromagnetic wavelength. In addition, we demonstrate the specular nature of the field scattered by such a surface. It is interesting to note that this is an excellent example of wave diffraction producing exactly the same scattered field as a perfectly flat reflecting surface once the point of observation is just a few tenths of the em wavelength above the crests of the rough surface. Computations also show that the power scattered into the specular direction is equal to the incident power, to within a few hundredths of a dB. Results will be presented for a surface where the technique fails to converge indicating that our preconditioning must be applied a second time to pre-sum some of the back and forth multiple scatterings on the surface before starting the iterative solution.

*"Computer Aided Approaches For Nanophotonics Research,"* Tae-Woo Lee, Louisiana State University. Over the past decade, venturing light manipulation in subwavelength scale has been accompanied by the advent of nanoscale probing and material processing techniques. First-principles computational modeling tools can help provide invaluable insights about wave phenomena inherent in photonic nanosystems. They can also be used to conduct inexpensive feasibility studies of a proposed device concept and to optimize a device design prior to fabrication. In this direction, a general concept of computer aided research and development (CARD) in nanophotonics will be pointed out. We will discuss how CARD can be systematically structured to facilitate nanophotonics researches. Rigorous numerical analyses from high performance computing resources and strong collaborative works are main frames of CARD environments. Next, we will present CARD enabled research outcomes as examples which indicate significance of CARD in nanophotonics researches: 1) quasi-3D plasmonic crystals for bio-sensing application, and 2) coherent excitations of local plasmons for encoding/decoding optical signals. A newly established computing center, CCT, at the Louisiana State University will be briefly introduced which can serve as an ideal base station for CARD environments.

*"First Steps and Results with Quantum Many-Body Math,"* J.H. Eberly, University of Rochester. Many-party interactions that preserve or partially preserve non-local coherences such as quantum entanglement have the potential for applications going far beyond what is available classically. This is well known. What is still not widely appreciated is that such interactions also permit applications going beyond the boundaries of the part of quantum theory known as wave mechanics. In a nutshell, entanglement is more than duality. To see the implications of entanglement physics the principles of many-body mathematics must be reduced to calculable examples, and we will report on this process. Two-particle calculations using many-body mathematics have been completed. Among the new results are several predictions of evolution of the degree of two-body entanglement that is not smooth in time, even under conditions of smooth evolution of the constituent one-body density matrices. Experimental testing of this exotic behavior is currently underway.

## 5.2 PHOTONICS BREAKOUT DISCUSSIONS

On Tuesday morning Dowling led the photonics breakout discussions. It was decided to take the three primary goals of the QMHP thrust, as defined by the NSF, and to break these down further into photonics related discussion points. The primary goals are: (1.) High-performance computing devices and systems; (2.) modeling and simulation for photonic systems; (3.) quantum many-body principles devices and systems.

*High-Performance Computing Devices and Systems*

*Plasmonics:* The control of light at the nanoscale requires high-performance computing for the description of plasmonic systems. Plasmonics and the related field of nanophotonics rely on novel light-matter interactions, which permit the manipulation of the optical field in nanodevices. New analysis and visualization methods and approaches are clearly needed for this field. Such developments will be important to explore the interface of electronics with photonics. Quanta versus classical versions of the systems under study need to be identified and understood. A hybrid quantum classical theory is needed, since while the basic picture is well understood, it needs to be extended to account for interfacing of light with quantum systems. Noise from Landau damping must be studies. Analysis and understanding of the photonic and electronic detection processes needs to be done in this context.

*Quantum Computers:* As quantum computers become available it is important to be on the lookout for ways to exploit the hyper-parallelism of the quantum states as a new paradigm in high-performance computing. Beyond classical computing, simple quantum computers will have their primary first practical role as simulators of simple quantum systems. This aspect then one could envision simple quantum computers as a tool for designing better classical computers that operate robustly in spite of quantum effects, thus pushing Moore's law further. For example, once transistors reach single atom size, quantum effects will dominate. However by simulating arrays of such transistors on a simple quantum computer, we may find ways to make it behave classically for standard classical computation, or force the quantum behavior to develop even more powerful quantum computers. The understanding of the role of entanglement, decoherence, and other such quantum effects in quantum-sized machines will become increasingly important.

*Optimization and Inversion:* Electromagnetic inverse problems always typically run a forward solution and the main issue is that, in general, inverse algorithms are slow and difficult. There is need to introduce new tools and standardize them. The Human interface is very important both in providing educated guess and in interpreting the final result. Integrate adaptive methods and non-orthogonal meshes. Any progress in inversion protocols that allow for electromagnetic devices with well-characterized designs will greatly advance the field of quantum photonics.

*Adaptive Computing:* New algorithms need to be designed to analyze multi-scale and multi-physics (photons, electrons and photons). The implementation of an efficient feedback loop in the simulation algorithm is critical. Novel numerical framework for quantum system designs, such as the path-integral approach, need to be considered. In this context quantum trajectory and Monte Carlo simulation techniques from quantum optics will be important.

*Quantum-Classical Boundary:* The development of a strategy for approaching classical quantum interface issues becomes important when designing quantum devices at the nanoscale. Which models for the electromagnetic field should be used and for which systems? Quantum computing and classical computing issues and understanding their roles become important. Post-processing and the role of "projection problem" in quantum computing based design and simulation needs to be understood. Is it possible to design tests to decide which regime the system is in — quantum or classical?

*Materials Aspects:* High performance computing design of NV centers in diamond will be important for the development of room-temperature quantum devices for use in quantum information processing. Growth approaches: bottom up versus top down need to be considered and modeled appropriately. One needs to understand how to develop quantum-entangled materials by design. A hierarchy of materials for quantum photonic devices needs to be considered for modeling purposes. Again the inverse problem approach to design new materials needs to be considered.

*Scale-Up and Integration:* Unforeseen problems are introduced by large-scale integration of photonic devices. The whole is, to first order, never just the sum of parts. We need to understand how to go from atoms and photons to macro-sized device and associated hardware. Examples are cavity quantum electrodynamics and quantum optics in fibers. Modular computational boxes and behavior at interfaces will be important to understand. Do the rules of quantum superposition apply to the realm of self-assembly, protein folding, etc.

### 5.2.1 Modeling and Simulations for Electronic and Photonic Systems

*Quantum Entanglement:* QMHP need to investigate new designer materials employing entanglement. A difficulty for such a goal is how to visualize entanglement for simulations and devices. We need to understand what are the bounds on the entanglement effects and how can we quantify the degree of entanglement. Also we need to educate the quantum electronics community that entanglement plays a role in a vast host of novel devices, not just in quantum computing. For example in the photonics realm it is now understood that, in addition to quantum computing and quantum communications, it is now understood that entanglement is a resource for a number of emerging quantum technologies such at quantum imaging, quantum metrology, and quantum sensing — all topics of recent Department of Defense thrust areas. Entanglement and quantum computing are not synonymous, and the role of entanglement in burgeoning new technologies is just beginning to be understood.

*Nonlinear Effects:* Efficient simulations of the light-matter interactions for optically nonlinear materials are needed. Such simulations are often very sensitive to phase errors. Classical computation issues also become important at the nanoscale at the energy expended per bit decreases. Energy constraints may play a role in the limitations for future computational advances, such as limits in computation speed because of the power. Designer nonlinear optical materials would benefit from new understanding and computational approaches for high-performance-based optical materials design approaches.

*Phonon Control:* For the design of optical devices at the nanoscale it will be more and more important to understand the role of photon-phonon interactions, which are typically viewed as a source of noise and decoherence. However it may be possible to control and even exploit the phononic nature of matter at these levels. Can we harvest phonon energy, say through some nonlinear Maxwell Angel (instead of the demon)? The answer to this question is very relevant for power control. Sideband phonon cooling is an example, as well as quantum photonic devices that mimic similar devices in the photon world.

*Photonic Crystals and Optical Metamaterials:* Recent advances in these fields in the past 20 years have shown that, to a great degree, the photonic properties of matter can be engineered by the careful design of optical materials whose properties are inherited from an interplay between their constituent atoms and their geometry. We have seen applications of photonic crystals to thermal control and improved thermophotovoltaics, and applications of metamaterials to such exciting devices as an invisibility cloak. Most of this work critically depends on the design of novel photonic devices with high-performance computers, and novel algorithms and inversion tools are sorely need. Much of the work in the past 20 years has treated the light field classically. However, as the structures continue to decrease in size, the quantum nature of the light field will play an ever more important role, just as in electronics the wave nature of the electron becomes important at small scales. At the single photon level, photonic crystals and metamaterials will have important application to quantum information processing, such as to optical quantum repeaters and all optical quantum computers. Optical computing is inherently parallel computing, and novel photonic devices may see the resurrection of search for the niche of the optical computer.

*Distributed and Grid Computing:* As in any thrust area, high-performance computing more often than not now means distributed and grid computing. For particularly difficult photonics design and inversion problems, 3D photonic crystals and metamaterials for example, any one institution is unlikely to have the requisite computational power locally to make headway. The NSF initiatives in grid computing and particularly allowing smaller universities to access the grid in a user-friendly way will be paramount. New numerical methods that run on such grids, such as pre-conditioning and renormalization of the problems, will be important, as well as method that go beyond the perturbation approach, such as is required for the modeling of optically nonlinear systems.

### 5.2.2 Quantum Many Body Principles Devices and Systems

*Entanglement:* Again we want to point out that the terms "entanglement" and "quantum information" are not synonymous, at least not in the quantum optics community, which is investigating applications of photon and atom entanglement to not only quantum information processing but to quantum imaging, quantum metrology, and remote sensing. Critical is the property of some entangled states to show super-resolution below the Rayleigh diffraction limit and super-sensitivity below the shot-noise limit. How to quantify entanglement in terms of measurement, such that quantum information community can talk to other communities will be an important endeavor. Developing better connection between entanglement measures and their experiment implementations is crucial. Extension of the entanglement for larger number of qubits is important, and given that quantum computers do not exist yet perhaps high-performance classical computing simulations of quantum entangled systems can take us part of the way. A classical computer cannot efficiently simulate a quantum computer, but making the slow classical simulations better and running them in supercomputers will take us some of the way.

Light-Matter Experimental Issues: In the quantum optics realm, the issues of light-matter interactions, particularly in optically nonlinear systems and in system design by inversion, require ever more powerful computing resources. Advances in the field can then draw a lot from advances in high-performance and grid computing. Multi-physics, or the physics at multiple length, energy, and time scales is particularly challenging from the computational perspective, and inroads in the QMHP thrust could help tremendously there. For example, the design and optimization of single photon sources, processors, and detectors is a computationally intense set of interrelated tasks, which could lead to breakthroughs in the fields of all optical quantum information processing, as well as quantum optical imaging, metrology, and sensing. Here then a close role of the models with experiment is required to make sure the models correspond to realistic situations.

*True Quantum Many-Body Theory:* Entanglement models, beyond perturbation models, such as solutions to the 3D Ising model are needed. Spin lattice models offer an interesting interdisciplinary area where quantum phase transitions, entanglement characterization, and perhaps quantum computing all meet. Density functional applications to spin lattice systems, as well as renormalization group theoretical methods will help make such systems computationally tractable. Taking the condensed matter theory beyond wave mechanics and reformulating condensed matter theory in new terms that invoke entanglement will be an important research area. How quantum many body theory goes into modeling large devices will need to be understood. Such modeling depends on the functionality of the device, and perhaps is it possible to take a black-box approach to large quantum systems designs. For example, how do we go from our understanding of the behavior of single dots to an understanding of arrays of millions of interacting quantum dots? Design of the macro-system to minimize deleterious effects, while exploiting knowledge of the micro-system will be paramount. Micro-macro packaging and a grand picture needs to be evolved.

## 5.3 PHOTONICS PANEL DISCUSSION

*Chair:* Jonathan Dowling

*Panelists:* Robert Trew (NCSU), Marko Loncar (Harvard), Lu Sham (UCSD), Joseph Eberly (University of Rochester)

*Robert Trew:* New thrust areas need to tie the NSF program to real applications and to fit in the big picture of things. Need to attack the fundamental problems of energy conversion and phonon control. We need also to attack new problems. What is beyond Moore's Law? Nothing proposed so far for a post-Moore's law paradigm has proven to work. For example, there is nothing shown to work, practically and economically, that could replace the CMOS-based technology. What is the big problem that you can try to solve if funded by the NSF QMHP program? One needs also to create a measure of success in basic research. Exceptions to this rule are for case studies such as GPS. Who would have predicted that work in molecular spectroscopy would give rise to GPS? To conclude, the QMHP program needs to fit in the big picture.

*Marko Loncar:* Connecting high-performance computing to experimental developments is important. In nanophotonics and photonic crystals rapid progress on the experimental side needs to interact with high-performance computing side. Boundaries between physical systems, such as plasmons interacting with atoms, need to be explored. Address the issue, for example, of why specific NV center diamond samples work for quantum information processing and others do not. Also a boundary issue: we need to understand how quantum dots coupling to external environment. It will be important to address problems of all-optical switches in silicon. Is it possible to engineer photon properties to cool quantum dots? An analysis of the feedback in photonic systems at very low power levels needs to be done.

*Lu Sham:* How to quantify the entanglement in quantum systems? Bell's inequality can provide the way to test them, but are there any other ways. How can we best implement simulations in plasmonics, photonic, electronic systems? We need to create a standard for extracting the most relevant aspects from the classical and quantum interface. QMHP must address the problem of multi-physics and the problem of correlations between electrons, photons, and phonons. Entanglement is a resource, but we need to develop the tools to quantify it, both in extending it to multi-particle entanglement and developing scalable experimental ways to measure it.

Joseph Eberly: Is there any relevant new idea for QMHP to consider? But perhaps that is not the right question. And the example above, with molecular spectroscopy and GPS, demonstrates that new ideas may be impossible to recognize before hand. Ideas in the area coherent light matter interaction in diverse systems, where photonics is in fact electronics: quantum dots SQUIDS, etc., are all devices that behave like quantum electrodynamic systems. These areas will provide developments both in terms of research relevance and applications. The applications that will come from quantum mechanics are not fully mature, and there is no measure of progress, but we need to distinguish what directions need to be pursued. To conclude, the QMHP program needs to contribute in smaller or larger steps to the evolution of science and technology.

*Discussions:* When the ideas are exciting, the people are excited and they invest theoretical, numerical, and experimental resources if the funding is there. An additional discussion revolved around the perception of the quantum electronics folks that all the talk about entanglement was synonymous with quantum computing. This notion was a surprise to the photonics folks, who think of quantum entanglement as a resource for an entire portfolio of emerging quantum technologies, including quantum computing, quantum communications, quantum metrology, quantum imaging, and quantum sensing. It was also asked why there was not much discussion of photonic crystals. The panelists and the chair felt that photonic crystals were are mature technology, with the principles of modeling them well understood, but where high-performance computing would pay off in the long run in terms of structure design and, as Eli Yablonovitch calls it, the inversion problem. In addition new photonic technologies, such as metamaterials, and nanophotonic and plasmonic devices were much in a much early stage of development and here the basic theory and understanding of the properties at the light matter interface were less well characterized and would be a very fruitful area for QMHP development.

# 6  APPENDIX

The appendix consists of input from individual attendees who were invited to contribute directly to the report.

*QUANTUM ELECTRONICS*

**Susan Coppersmith, University of Wisconsin**

The challenge of doing quantitatively accurate device and systems modeling is an enormous one because the feature sizes in current and future devices are small enough that quantum effects are critical, but the overall devices and systems involve enormous numbers of components.  The critical need for enhanced fundamental understanding is underscored by the fact that the present basic strategies for electronics fabrication will soon face limits induced by fundamental physical properties (e.g., oxide thicknesses and feature sizes reaching atomic sizes), so that future performance improvements will rely on developing and perfecting novel materials systems and qualitatively new fabrication strategies.  The dominance of quantum effects on atomic length scales gives rise to the possibility of harnessing purely quantum effects such as entanglement for technological applications.  Developing robust and accurate computational modeling tools that incorporate the essential quantum-mechanical properties at a microscopic level into a system while incorporating the huge number of devices in a technologically relevant system is a key enabler for designing and optimizing future generations of electronic devices.

A fundamental property of quantum systems that gives rise to the intrinsic difficulty of accurately modeling the microscopic properties of devices is that the number of parameters needed to specify a quantum-mechanical state of N particles grows exponentially with N.  In contrast, a classical description of a system of N particles requires specifying a number of parameters that grows linearly with N.  Therefore, full quantum-mechanical treatment of all but extremely small systems is prohibitively computationally expensive, and so it is vital to incorporate analytic insights and multiscale techniques to be able to do quantitatively accurate modeling of complex devices and systems.

*QUANTUM PHOTONICS*

**Gary Brown, Virginia Tech**

<u>Modeling and Simulation</u>

   Modeling and simulation are two very important elements for understanding, predicting, and designing new photonic devices and systems. They take on an exceptionally significant role in situations where there is an interface between fairly well understood science and science, which has elements that are still emerging, i.e., such as electromagnetics as compared to certain aspects of quantum mechanics. Both of these two disciplines are important to the development and fielding of future new and revolutionary photonic devices and they form the core of any associated modeling and simulation efforts. It is absolutely essential that accurate, robust, and versatile computer implementations, or "solvers", for both electromagnetics and quantum mechanics must be available and ready for use by physicists and engineers. There are a number of criteria that such software must satisfy. Among these are (1) accurate modeling of electromagnetic and quantum mechanics aspects of the boundary value geometry representing desired devices, (2) efficient and rapid computational predictions of device performance, (3) broad capabilities in so far as adaptability to geometrical and material changes.

Among some of the attributes that would be desirable of such device-focused software are the following;
- the use of renormalization or other partial summation methods to overcome the limitations of perturbation and asymptotic analysis,
- the ability of the software to be scaled up to larger geometries, differing materials, and/or multiple configurations,
- the assurance that the classic limit of quantum mechanics predictions will be attained in the appropriate dimensional/material limit, and
- the combining of classical electromagnetics and quantum mechanics in such a way as to provide a near optimal computing engine for efficient and comprehensive simulations.

Renormalization or partial summation is one way of recasting the basic equations of interactions into forms, which extend well beyond perturbation techniques and are amenable to numerical methods. Such approaches usually also have the advantage of giving more insight into the physical interactions that take place on all scales. They also may yield insight into the capabilities and limitations of "effective parameter" approximations that have been used previously to understand these classes of problems.

The aspect of scaling revolves around the assumption that the whole is predictable given the individual parts; such an assumption is critical to scaling the size of the device simulation to larger configurations. It may be too optimistic to require that the "whole is equal to the sum of the parts" for very few numerical problems scale in such a manner; however, predictability is absolutely essential. It is also very important that computations and simulations on the level of atoms and photons can be translated into results at the macro-device level. One of the desired products of such simulations is, in the long range, self-assembly where systems of devices are capable of rearranging their configurations in such a way as to be adaptable to a change in the desired simulation.

**Joseph Eberly, University of Rochester**

(I) The device domain where quantum principles affect photonic-electronic operating principles has already begun to be entered. Prototypes of photonic-electronic devices have begun to be conceived in such ways that they can perform tasks relevant for communication, storage and data manipulation in close analogy to, and with many of the same advantages as, what is projected for cavity QED devices, and with some advantages over them. Clean photonic-electronic interfaces now require quantum correlation theory for quantitative calculation, but such a quantum theory is still in a primitive state. Significant advances in quantum transport, quantum relaxation, and many-body spin decoherence theory will be required. One avenue to proceed was described, via Kraus operator many-body quantum mathematics. Also, phononic applications appear to have intriguing futures, including single-phonon detection via coupling to either linear or nonlinear nanomechanical-electrical system (NEMS) devices. This is a future direction that should be targeted. One example of the inroads being made into this territory was reported in connection with theoretical and experimental progress in calculating and observing so-called entanglement sudden death [J.H. Eberly and Ting Yu, Science (April 27, 2007)]. Serious innovation in quantum modeling is needed to accompany these thrusts, and there was substantial agreement that the meaning of "true quantum behavior" is poorly defined. The boundaries of the areas where the influence of quantum entanglement might be felt need to be explored. It was accepted that the potential for substantial progress in fresh directions is very great in these areas, and the need to better understand the quantum-classical interface was mentioned more than once. New modeling directed to efficient implementation of path-integral and semiclassical quantum trajectory approaches to both dynamic and decoherent processes was endorsed as a key goal for the near future.

(II) Two days of QMHP discussions revealed a number of interesting projects under way. Current ideas are being pursued in both the electronic and photonic domains of QMHP. Photonic crystal design has serious topics to challenge it. Energy-harvesting ideas are being pursued (but were not clearly justified). High-performance computing is a goal, but some Workshop participants doubted its ultimate usefulness in a real-life engineering environment. Innovations that engage many-quantum coherence in a new way were generally endorsed for their potential in spin, photonic, phononic, and plasmonic device applications. They can be the basis of completely new photonic and electronic benefits flowing from advanced engineering research all the way to empowering the second billion. It was accepted by essentially all participants in the Workshop-ending Panel that what is truly new can't be expected to have immediate relevance to devices, and what is sure to be device-relevant shouldn't be expected to be fundamentally new. Both fundamentally new projects and clearly device-relevant projects need support if the QMHP Program is to have long-term viability.

**Taewoo Lee, Louisiana State University**

QMHP workshop report – nanophotonics

The capacity of numerical simulations has now expanded largely due to the advent of high performance computing (HPC) techniques. The significance of HPC-enabled simulations is that one can apply rigorous numerical schemes to solve physical problems in a realistic full-size scale. In general, this kind of numerical simulation is useful when interesting physical systems are complex and intuitively unpredictable, where one should not simplify or approximate the system of interest. It would be more desirable if HPC-based research activities could involve scientific subjects that might open up new possibilities for future device applications, and pioneering engineering subjects that might imply new device applications from the previous scientific research outcomes.

Among many promising areas in nanophotonics, surface plasmons (SPs) and relevant systems would be proper subjects of research 1) for which HPC-based simulations can be very useful and 2) from which a person's creativity can be examined in terms of novel device ideas. In spite of their long history, it has been only in recent years that SPs have been seriously considered as a research subject. Due to its large potential to achieve light manipulations in nanoscale, there has been explosive attention paid to this subject during this short period of time. Furthermore, strong local field intensities in SPs can also be useful to boost up optical nonlinear effects that also bring new possibilities for realizing nonlinear photonic devices. If researches are aimed at exploring SP systems to discover desirable features for future device applications, one should employ rigorous numerical tools so that any unexpected physical events are not overlooked. Hence, HPC-enabled simulations would be a great help for this type of research. However, there are still challenges in HPC simulations to explore SPs: for example, incorporating nonlinear effects in SP interactions, describing quantum mechanical effects in case of single or a few countable photon interactions in SP systems, etc. Therefore, it is desirable to investigate for algorithm developments based on HPC approaches.

In summary, a new research program would be best designed to encourage scientific communities to examine new possibility for devices or system applications. To be distinguished from core engineering research, scientific studies need to be emphasized as bringing foundations of new device concepts. SPs would be one of choices in nanophotonics, but there could be many other choices as long as they are new and promising. Developments of new numerical techniques may also be necessary for targeted nanophotonic problems. In all cases, HPC simulations should be base tools for pursuing those research activities.

**Ting S. Luk and Ihab El-Kady, Sandia Laboratories**

Energy scavenging and mobile energy sources

One of the most fundamentally important and strategically relevant directions of research, from a national security point of view, is energy independence. To this end, the major thrusts were focused on the search for new and alternate renewable sources of energy, along side the attempt to enhance device operational efficiency and hence reduce power consumption. In particular two energy sources still require large fundamental scientific advances in order to ensure a full realization of their potential, namely: solar and thermal energies.

Most attempts to harness solar energy have focused on the material development of photovoltaic (PV) cells. These however have come to an unfortunate abrupt stall as the efficiency of the PV cell is capped by the one-to-one photon to extractable exciton conversion economy on top of the Shockley-Queisser limit, which severely limits the efficiency of PV devices.

A promising avenue was recently demonstrated where it was shown that in IV-VI quantum dot systems, an energetic photon can indeed generate multiple-excitons, and that indeed under certain conditions [1] this multi-exciton scheme would result in the enhancement of solar-PV systems [2]. However a major impediment is inability to extract these multiple excitons before the non-radiative processes prevail as a result of the Coulomb blockade restrictions that prevent the extraction of a single exciton at a time. A promising scheme that remains unexplored is the use of photonic crystals to enhance the radiative decay lifetimes of these quantum dot systems to the point where they can compete with the non-radiative decay processes, thus allowing for the extraction of the exciton energy in the form of photons which can intern be matched to the band gap of a PV cell hence avoiding at the same time Shockley Queisser issues.

On the other hand, one naturally abundant source of energy that is routinely overlooked is thermal energy. Viewed usually as exhaust or waste, thermal energy is regarded as the unfortunate life cycle end of what used to be useful energy. An alternate way of looking at the problem is to try to find ways of scavenging this conceptual "waste" energy by recycling parts of it into the more useful form of electrical energy.

One possible way of avoiding this scenario is allowing for non-equilibrium conditions to prevail in the emitter side and as such break through the usual Plank distribution law cap on thermal energy extraction.

R. DiMatteo demonstrated a promising avenue, is tapping into the evanescent field energy that is by definition a non-equilibrium thermodynamic system [3]. There the initial modeling of Micron-gap Thermophotovoltaics (MTPV) has shown significant performance improvements relative to conventional far field TPV. These included up to a 10× increase in power density, and 30-35% fractional increase in conversion efficiency. However the prevailing drawback was in achieving the desired micro gap spacing and maintaining it in practical harsh environmental conditions.

However, recent studies of a number of polar semiconductor and dielectric materials that support surface phonon polaritons show that these materials exhibit quasimonochromatic thermal emission over the entire range of temperature 300-600K and the distance from the surface of up to 10μm [4].

Despite the anticipated extension of the evanescent to several microns, this remains a very impractical scenario from the application side of view as such small distances are not only difficult to maintain mechanically, but also the close proximity of the PV cell to the emitter will inevitably result in the subsequent heating of the cell and thus result in efficiency reduction. A plausible solution may rely on finding the correct surface texturing of the polar emitter to create long lived surface states with long spatial extensions, one such possibility can be photonic lattices. However the development of such solutions remains highly difficult, as it requires care full engineering of the complex photonic band-diagrams and bandedge matching to the PV cell. While such ideas may not be too far fetched they require a grater fundamental understanding of the inner workings of photonic crystals and especially of the more complex higher order bands.

A final plausible scenario is to view phonons with the same footing as phonons. In other words, to view them as quantum mechanical carriers rather than destroyers of energy, and to try to harvest phonons in the same way photons are routinely harvested. One such possibility can be established through the construction of a homomorphism between photonic crystals and phononic crystals. Furthermore, recent developments of phononic crystal fabrication and modeling of phononic crystals have revealed the possibility of reaching GHz and THz frequencies [5-6]. This opens the door to ambient energy harvesting. In addition, the ability to mold and control phonon flow can immediately impact device operational efficiency, where our ability to rapidly cool a device by ballistic phonon propagation, or the creation of a phonon shielded to reduce the signal-to-noise ratio or even to manipulate the heat capacity of a device will directly reduce the energy running coast of any given device.

The ideas presented above, are evidently in the early stages of development and demand a large amount of fundamental understanding of the inner workings of photonic and phononic devices, and though a path may be laid out, significant fundamental scientific effort has to be devoted to achieving a better understanding of the mechanics of energy loss and recycling of phonons. Most importantly however, the ideas presented above lay the foundation for mobile energy sources and self powered device packages that rely solely on energy scavenging and recycling.

1. Richard D. Schaller, Milan Sykora, Jeffrey M. Pietryga, and Victor I. Klimov, NANO LETTERS, 6(3), 424 (2006).
2. Victor I. Klimov, APPLIED PHYSICS LETTERS 89, 123118 (2006)
3. R. DiMatteo, et al, THERMOPHOTOVOLTAIC GENERATION OF ELECTRICITY: Sixth Conference on Thermophotovoltaic Generation of Electricity: TPV6. AIP Conference Proceedings, Volume 738, pp. 42-51 (2004).
4. V. N. Sokolov, B. D. Kong, K. W. Kim, and R. J. Trew, APPLIED PHYSICS LETTERS 90, 113106 (2007)
5. I. El-Kady, R. Olsson, APPLIED PHYSICS LETTERS in review.
6. I. El-Kady, R. Olsson, PHOTONIC NANO STRUCTURES FUNDAMENTALS AND APPLICATIONS, to be published

**Lu Sham, University of California San Diego**

Device Modeling and Simulation in the Regime between Classical and Quantum

The size of the electronic devices has been reduced to the extent that the limit will soon be reached where existing functionalities depending on the use of classical quantities of charges become unreliable. Quantum effects will affect the functions of yet smaller devices and methods of combating them are being considered. On the other hand, single particle properties are being explored for quantum machines, such as information processing and computing. In between these classical and quantum regimes lies an intermediate regime of opportunities to harness the quantum effects of multiple particles for device applications of superior performance. Glimpses of such promises already exist in the measurement protocols with accuracy beyond the standard quantum limit [1] and quantum lithography method below the diffraction limit [2]. Physics in the border region of classical and quantum physics begins to be explored [3]. Its unfamiliarity and difficulty demand greater innovation in theory where numerical simulation may well be the most effective tool.

In CMOS, the vertical confinement of an MOS component is in the quantum region and is successfully accounted for. The horizontal confinement lateral to the current channel is currently being studied in quantum wires. It is the effort to reduce the channel length below 20 nm, which will bring new quantum effects creating problems for defining the information bit by the potential in the capacitor and for the ballistic nature of transport in the channel for gate operations. This is an example of the borderline regime. Similarly, in further development of nanophotonics, the number of photons in a functional pulse will be smaller so that shot noise becomes dominant. But we can work to make quantum effects beneficial. Entanglement could be built with an N photon state and a zero photon state along different paths [4]. Interferometry [1,5] to beat the shot noise limit making use the N, 0 entanglement could become routine engineering. Single photons could serve as the classical communication bits, a feat intermediate in difficulty level to implement between nanophotonics and the quantum bit of linear combinations of zero and one photon states.

An example of an integrated optoelectronics system might be borrowed from the quantum network of atom-cavity quantum electrodynamics [6]. The difference would be classical bits in the node for electronics and laser control in cavity with a nano-optical pulse rather than with a single photon qubit [7]. The optical communication channel could be either a wave-guide or a fiber. The entire system could be integrated in a chip of photonic lattices [8]. The wave-guide may serve as an entropy dump for cooling [9]. The Raman control of the quantum dots may also be used to convert phonon energy to amplify photon energy by the working principle of side-band cooling pioneered in trapped ions. The size of the nano-cavity could provide pioneering work in electromagnetic field simulation in the border region between waves and photons.

The criterion for a system to be in the border region might be the Bell inequality test. A strategy needs to be developed for the theory in the border region, exploring questions such as the proper mixed use of classical and quantum methods or extension of current semiclassical methods, including development of numerical solutions of the path integral method. For particle numbers ranging from mesoscopic to a few, the many-body problem is of great interest, encompassing the techniques from field theory and condensed matter physics to finite nuclei theory. Yet, because of the engineering emphasis on control, questions to be asked such as entanglement resource require a fresh approach. For applications, it is clear from the recent discovery of "sudden entanglement death" of two qubit entanglement in an open system [10] that there is much to be done. Principal among the issues are a measure of entanglement, which can be measured directly, and the investigation of multi-partite entanglement.

This program will provide opportunity for very basic development of quantum theory to enable effective treatment of systems in the borderline regime between quantum and classical and for pioneering modeling and simulation in systems which are next in the natural development of the engineering field. The benefits to the field are new theories and understanding and means of control of these in-between sized systems. The effort may well start a revolution in methodology for this intermediate regime and, furthermore, pave the way for major breakthroughs for engineering in the quantum limit. The integrated opto-spintronics system described above could provide a blue print for a pre-quantum computer, for a hybrid classical and quantum communication system, and for converting waste heat in the form of lattice vibration energies into light energy.

An extremely important mission of this program is the quantum education of future engineers who will require insight into quantum mechanics as well as integrating the information theoretic method in quantum engineering. The tradition method of teaching quantum mechanics would not suffice because of the aspects of control and environment impact in engineering applications to quantum machines. This presents a great challenge. If successful, it would fulfill an important societal need.

**Robert Trew, North Carolina State University**

Surface excitation and its manipulation in the near field is an important and timely subject of investigation. Specifically, the interplay of the surface excitations with the electromagnetic radiation through various surface structures (such as grating) is known to induce phenomena beyond the conventional diffraction limit. Revolutionary developments are expected with potential applications ranging from sub-wavelength imaging to THz generation, energy harvest, heat removal, etc. Detailed physical understanding of the electromagnetic wave properties including the evanescent field at the nanoscale distances is required to take full advantage of this development.

**Eli Yablonovitch, UCLA**

The Application of New Algorithms to Solve Inverse Problems in Engineering

The solution of Maxwell's Equations is central to many areas of technology. I will give the following real-world business examples in which a more accurate Electromagnetic computation redounds right down to the bottom line of a business:
1. Cellphone Internal Antenna design
2. Nano-photonic circuits in Silicon-on-Insulator wafers
3. Design of photo-masks for Photo-Lithography
Simulation alone is insufficient in all these real life examples. Ideally the computation should encompass an inverse design algorithm that actually solves the engineering challenge.

The challenge for NSF is to anticipate the science and technology of the future. Indeed this is an obligation for all of us in the scientific fields. One of the greatest influences on societies around the world has been Moore's Law of the progression of electronics and communications. Moore's Law has been based on miniaturization, but miniaturization is coming to a culmination. Devices will soon be as small as individual molecules.

One of the biggest questions is what new challenge will be of equal importance, in its influence on society, as the progressive miniaturization of technology? That new challenge should be power dissipation, or energy per bit function. The natural unit for this is kT, thermal energy, per process one bit of information. Currently this requires on the order of $40 \times 10^6$ kT of energy for simple functions such as communicating one bit of information a distance of a few microns on chip. While this may improve somewhat based on further levels of miniaturization, it is expected that the energy per unit function will still fall a factor 1000 short, even at the final culmination of miniaturization.

The idea here is not to conserve the relatively modest amount of energy used in information processing. This is not about saving ergs. Rather the goal would be to answer how would the world change if we could do one million times more information processing for the same size battery in a portable application? Or how would one million times more processing change the world of highly parallellized fixed base computation, simulation, and inverse design? Thus energy per bit function becomes the new parameter in this second Moore's Law.

Regarding what new functions would be enabled, there are many challenges, but applied mathematics & engineering are emerging as some of the greatest beneficiaries of both the first and second Moore's Law. Engineering is about design, and there are many design problems that have always been the province of the skilled and highly intuitive engineer. Moore's Law has enabled tremendous progress in simulation, but now it is emerging as a method of solving inverse design problems. In this respect computational power could substitute for an intuitive engineer, and represent a form of artificial intelligence.

Among problems of this type, there are a number of electromagnetic design issues that are of great commercial importance. Likewise there are mechanical problems where inverse design can be paramount. On the other hand quantum problems, while challenging don't necessarily lead to actionable results. In quantum materials science, there are few levers available to designers; change the atomic number? Therefore the most immediate emphasis needs to be to identify and encourage the solution of major classical inverse algorithms that could have immediate impact on engineering.